\documentclass[prl,amssymb,twocolumn,tightenlines,showpacs,superscriptaddress]{revtex4}

\usepackage{graphicx}
\usepackage{amsmath}

\begin{document}

\title{Bunching of Bell states}

\author{H.S. Eisenberg}
\affiliation{Department of Physics, University of California, Santa Barbara, California 93106, USA}
\affiliation{Racah Institute of Physics, Hebrew University of Jerusalem, Jerusalem 91904, Israel}

\author{J.F. Hodelin}
\affiliation{Department of Physics, University of California, Santa Barbara, California 93106, USA}

\author{G. Khoury}
\affiliation{Department of Physics, University of California, Santa Barbara, California 93106, USA}

\author{D. Bouwmeester}
\affiliation{Department of Physics, University of California, Santa Barbara, California 93106, USA}

\pacs{03.65.Ud, 42.50.Ar, 42.50.St}

\begin{abstract}
The bunching of two single photons on a beam-splitter is a
fundamental quantum effect, first observed by Hong, Ou and Mandel.
It is a unique interference effect that relies only on the photons'
indistinguishability and not on their relative phase. We generalize
this effect by demonstrating the bunching of two Bell states,
created in two passes of a nonlinear crystal, each composed of two
photons. When the two Bell states are indistinguishable, phase
insensitive destructive interference prevents the outcome of
four-fold coincidence between the four spatial-polarization modes.
For certain combinations of the two Bell states, we demonstrate the
opposite effect of anti-bunching. We relate this result to the
number of distinguishable modes in parametric down-conversion.
\end{abstract}

\maketitle

The bunching of photons is a manifestation of their bosonic nature.
Although resulting from quantum interference, it is insensitive to
the photons' relative phase. Bunching of two photons on a
beam-splitter was first demonstrated in the famous experiment by
Hong, Ou and Mandel (HOM)\cite{Hong}. Later, bunched photon states
were shown to be useful for quantum limited interferometric
measurements\cite{Holland} and for beating the classical diffraction
limit\cite{Boto}. The bunching of three photons was also
experimentally achieved\cite{Mitchell}. Photon bunching became a
fundamental tool in quantum optics. It is used in Bell state
analysis\cite{Michler}, teleportation\cite{Bouwmeester} and more.
Recently, the basic principles of bunching have also been found to
be useful for quantum information processing with linear
optics\cite{Knill}. An example is the nonlinear phase-shift on a
beam-splitter for the quantum controlled-NOT gate\cite{OBrien}.

Photons can bunch also in circumstances where there are no
beam-splitters. Using the equivalence between the operation of
beam-splitters on two spatial modes and the operation of waveplates
on two polarization modes\cite{Shih}, a horizontally polarized and a
vertically polarized photon can bunch when their polarization basis
is rotated. Polarization interference experiments are simpler as
they involve a smaller number of spatial modes.

The most common source for photon bunching experiments, as well as
for photon entanglement, is parametric down-conversion (PDC). As an
intense pump beam is passing through a crystal possessing
$\chi^{(2)}$ nonlinearity, some of its photons can split into two.
For type-II non-collinear PDC, these two photons are emitted into
two spatial modes (referred to as modes \textit{a} and \textit{b}),
and can exhibit polarization entanglement\cite{Kwiat}. It is helpful
to use pulsed pump sources due to their well defined timing.
Recently, many experiments have used configurations where a pump
pulse passes through the nonlinear crystal
twice\cite{Bouwmeester,Lamas-Linares}. In this Letter we show how
four photons that originate from such two passes avoid being equally
distributed between the four possible spatial-polarization modes by
phase-insensitive destructive interference\cite{Lim}. This result
has many similarities to the HOM bunching. Its main difference
though is that instead of the bunching of two single photons in Fock
states, the bunching occurs between two composite states -- the Bell
states. As opposed to previous double-pass experiments, the result
is unaffected by the amplitudes of two-pair emission in one of the
two passes. We also show how, unlike in the HOM case of single
photon states, for certain combinations of Bell states the bunching
transforms into anti-bunching.

Non-collinear type-II parametric down-conversion creates the
following bi-partite state in a single pass of the nonlinear
crystal\cite{Braunstein}
\begin{subequations}\label{Psi}
\begin{eqnarray}
|\psi\rangle&=&\frac{1}{\cosh^{2}\tau}\sum_{n=0}^\infty\sqrt{n+1}\tanh^n\tau|\psi^-_n\rangle\,,\label{Psia}\\
|\psi^-_n\rangle&=&\frac{1}{\sqrt{n+1}}\sum_{m=0}^n(\textendash
1)^m|n\textendash m,m\rangle_a|m,n\textendash
m\rangle_b\,,\label{Psib}
\end{eqnarray}
\end{subequations}
where $|m,n\rangle_i$ represents $m$ horizontally and $n$
vertically polarized photons in mode $i$. The magnitude of the
interaction parameter $\tau$ depends on the nonlinear
coefficient of the crystal, its length and the intensity of
the pump pulse. The one-pair term ($n$=1) is the familiar
$\psi^-$ Bell state\cite{Bell}. We concentrate on the case
when two indistinguishable photon pairs are produced ($n$=2).
This term contains three elements and it is written in the
above formalism as
\begin{equation}\label{Psi2}
|\psi_2^-\rangle=\frac{1}{\sqrt{3}}(|2,0\rangle_a|0,2\rangle_b-|1,1\rangle_a|1,1\rangle_b+|0,2\rangle_a|2,0\rangle_b)\,.
\end{equation}

Previously, it has been shown how a four-fold coincidence between
the four modes $a_h,a_v,b_h$ and $b_v$ is forbidden for the state of
Eq. \ref{Psi2}, when the photons of mode $a$ are rotated to an
orthogonal polarization basis compared to the photons of mode $b$
\cite{Banaszek,Tsujino,Eisenberg2}. For example, if mode $a$ is
observed at the linear horizontal-vertical ($hv$) basis and mode $b$
at the linear plus-minus ($pm$) $45^\circ$ basis (or the right-left
circular basis $rl$), an $a_ha_v$ coincidence collapses the state to
the middle term of Eq. \ref{Psi2}, and the polarization rotation
bunches the $b$ mode photons such that
$b_hb_v$$^{\underrightarrow{\lambda/2}} b_p^2-b_m^2$. Thus, a
four-fold coincidence of the form $a_ha_vb_pb_m$ can never occur.
The same result applies to any two orthogonal polarization bases due
to the rotational symmetry of Eq. \ref{Psi2}. Nevertheless, when
more than one mode are collected for each spatial-polarization mode
(as in a double-pass configuration), these four-fold coincidences
can be observed.

\begin{figure}[tbp]
\includegraphics[angle=0,width=86mm]{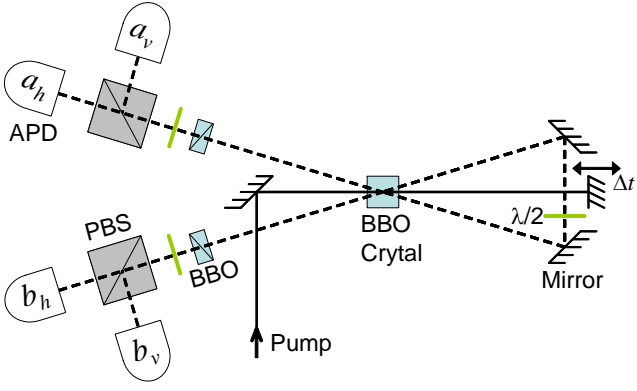}
\caption{\label{fig1}The experimental setup. The pump pulses pass
twice through the nonlinear BBO crystal, with a controlled delay
between the passes. Down-converted photons from the first pass are
re-injected into the crystal together with the pump second pass. Two
BBO crystals are used to compensate for temporal walk-off. Photons
are coupled through narrow bandpass filters into single-mode fibers
and detected by APDs.}
\end{figure}

In the experiment, we used a PDC source in a double-pass
configuration\cite{Lamas-Linares}. The nonlinear crystal is pumped
by 200\,fs pulses of 390\,nm wavelength from a doubled Ti:Sapphire
laser at a repetition rate of 80\,MHz. Down-converted photons from
the first pass are redirected into the crystal, to meet again with
the same pump pulse. The use of a double-pass configuration can
introduce distinguishability in a controlled way. As the pump back
reflection mirror (See Fig. \ref{fig1}) is translated, the timing
$\Delta t$ between down-converted photons from the two passes is
changed. This timing is an extra distinguishing quantum number that
can be varied continuously between turned on (by a delay longer than
the coherence length $t_c$) and off (at zero delay).

\begin{figure}[tbp]
\includegraphics[angle=0,width=86mm]{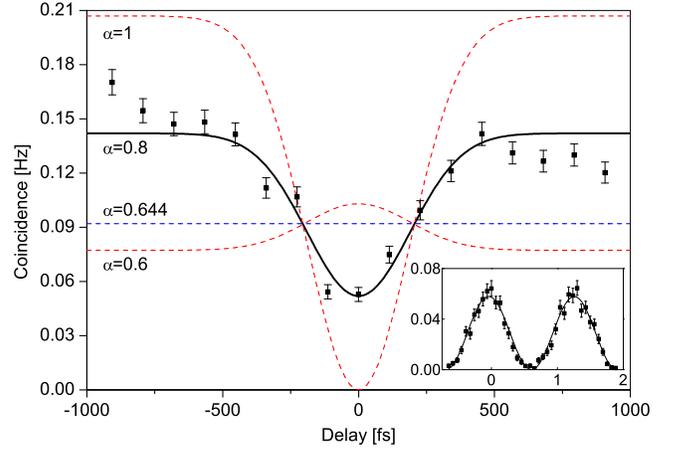}
\caption{\label{fig2} Measured four-fold coincidence rate (squares)
as a function of the delay $\Delta t$ between two $\psi^-$ states.
Mode \textit{a} is detected at the \textit{hv} basis and mode
\textit{b} is at \textit{pm}. Every point is integrated over 25
minutes. The curve for the best fit value of $\alpha=0.8$ is solid
while predictions for other $\alpha$ values are shown as dashed
lines. \textbf{inset}: a fine scan around zero delay reveals
oscillations.}
\end{figure}

For a pure $\psi_2^-$ state, no four-fold events can be detected at
orthogonal polarization bases and zero delay. Nevertheless, when
$\Delta t\gg t_c$, the temporal delay adds distinguishability
between the Bell states from the first and the second passes,
doubles the number of collected modes and gives rise to four-fold
coincidences. To demonstrate this, we define a ladder operator
$L^\dagger=(a_h^\dagger b_v^\dagger -a_v^\dagger
b_h^\dagger)/\sqrt{2}$, composed of creation operators of the four
PDC modes. The operator $L^\dagger$ creates a $\psi_1^-$ state when
applied once to the vacuum and a $\psi_2^-$ state when applied
twice. The four-photon state resulting from two passes of a pump
pulse in the down-conversion crystal with a phase difference of
$\omega \Delta t$ between them is \cite{Lamas-Linares}
\begin{eqnarray}\label{Psi2double}
|\psi_2\rangle &=&\frac{1}{4}(L_\textrm{I}^\dagger+e^{i\omega \Delta t} L_\textrm{II}^\dagger)^2|vac\rangle\\
\nonumber
&=&\frac{1}{4}({L_\textrm{I}^\dagger}^2+
2e^{i\omega \Delta t} L_\textrm{I}^\dagger L_\textrm{II}^\dagger+
e^{2i\omega \Delta t} {L_\textrm{II}^\dagger}^2)|vac\rangle\,.
\end{eqnarray}
The roman digits designate a distinguishing quantum number. The
distinction can be one of many options, such as the photon
wavelength, spatial mode or timing, as in this double-pass case. The
terms that result from the first and last operators in Eq.
\ref{Psi2double} do not contribute to the $a_ha_vb_pb_m$ four-fold
coincidence as they each create a $\psi_2^-$ state. Keeping the pass
number labels and discarding normalization, the resulting evenly
populated state terms, that would yield a four-fold event when
observing modes $a$ and $b$ at orthogonal polarization bases are
\begin{eqnarray}\label{Psi4fold}
|\psi_{4\textendash fold}\rangle=( a_{h,\textrm{I}}^\dagger
a_{v,\textrm{II}}^\dagger b_{p,\textrm{II}}^\dagger
b_{m,\textrm{I}}^\dagger&+&a_{h,\textrm{II}}^\dagger
a_{v,\textrm{I}}^\dagger b_{p,\textrm{I}}^\dagger
b_{m,\textrm{II}}^\dagger\\
\nonumber -a_{h,\textrm{I}}^\dagger a_{v,\textrm{II}}^\dagger
b_{p,\textrm{I}}^\dagger
b_{m,\textrm{II}}^\dagger&-&a_{h,\textrm{II}}^\dagger
a_{v,\textrm{I}}^\dagger b_{p,\textrm{II}}^\dagger
b_{m,\textrm{I}}^\dagger)|vac\rangle\,.
\end{eqnarray}
Because these terms originate only from the middle term
$L_\textrm{I}^\dagger L_\textrm{II}^\dagger$ of Eq.
\ref{Psi2double}, their amplitudes are neither sensitive to the
phase $\omega \Delta t$, nor to the amplitude balance between the
two passes. Like in the HOM bunching experiment, when the two input
states are indistinguishable, the pass indices are omitted and the
amplitude for $\psi_{4\textendash fold}$ disappears. However, the
absence of interference in the distinguishable case revives the
four-fold amplitude. As the delay $\Delta t$ is scanned the
distinguishability is varied, resulting in a dip for the four-fold
counts, centered at zero delay.

Figure \ref{fig2} shows the four-fold coincidence rate for
orthogonal polarization bases as a function of the delay between two
$\psi_1^-$ states. As predicted, the coincidence rate has a dip,
centered at the zero delay point. The dip width corresponds to the
coherence time of the two $\psi_1^-$ states\cite{Ou}.

Previously, the quality of the two-pairs state was defined by
$\alpha$, the probability for having the $\psi_2^-$ state, as
opposed to two distinguishable $\psi_1^-$
states\cite{Tsujino,Gisin}:
\begin{equation}\label{alpha}
|\psi\rangle=\sqrt{\alpha}|\psi_2^-\rangle+\sqrt{1-\alpha}|\psi_{1,\textrm{I}}^-\rangle\otimes|\psi_{1,\textrm{II}}^-\rangle\,.
\end{equation}
This model is problematic if we assume that it originated from two
modes as in Eq. \ref{Psi2double}. First, such a state has at least
60\% content of $\psi_2^-$, i.e. $\alpha$ has a non-zero minimal
value. Second, the $\psi_2^-$ content is the sum of two terms,
$\psi^-_{2,\textrm{I}}$ and $\psi^-_{2,\textrm{II}}$ from the two
modes (passes). Furthermore, two or more distinguishable modes are
possible even for a single pass. We would like to have a model with
arbitrary number of modes that will support results where $\alpha$
has lower values and explain the experimental imperfect interference
at the center of the dip. The multi-mode Hamiltonian and the
four-photon component of the state produced are
\begin{eqnarray}\label{Hmulti}
H&=&\frac{i\kappa}{n_d}\sum_{j=1}^{n_d} c_j e^{i\theta_j} L_j^\dagger +h.c.\,,\\
\label{Psi2multi} |\phi_2(n_d)\rangle &=&\frac{1}{\sqrt{C}}(\frac{1}{n_d}\sum_{j=1}^{n_d} c_j e^{i\theta_j} L_j^\dagger)^2|vac\rangle\,,
\end{eqnarray}
where $\kappa$ is a coupling constant that depends on the
nonlinearity of the crystal and the intensity of the pump pulse, and
$C$ is the proper four-photon component normalization. The
distribution of $c_j$ (real numbers) determine the number of modes
involved and their relative weight such that their average is
$\langle c\rangle=1$. As the operator $L^\dagger$ has two terms, the
state of Eq. \ref{Psi2multi} has $2n_d^2+n_d$ non-interfering terms
($3n_d$ originate from $(L_j^\dagger)^2$ terms and $2n_d(n_d-1)$
from $L_j^\dagger L_k^\dagger$), of which only $n_d^2-n_d$ can give
rise to the relevant four-fold events (half of the $L_j^\dagger
L_k^\dagger$ terms). For the simple case of equally weighted modes,
the intensity of $|\phi_2\rangle$ is $(2n_d+1)/(n_d^3)$, which is
higher for smaller $n_d$ values due to better stimulation. After
rotation to orthogonal polarization bases, four-fold events result
only half the time and for the equally weighted modes case their
probability per pump pulse scales as
\begin{equation}\label{rate}
P_4(n_d)\propto \frac{n_d-1}{2n_d^3}\,.
\end{equation}
When only one PDC mode is collected there are no such events, and
when the mode count increases, the rate peaks for $n_d=2$ and decays
for larger numbers. When the time delay $\Delta t$ is introduced in
the experiment, the mode number $n_d$ is doubled such that each term
from the first pass corresponds to a delayed term from the second
pass.
\begin{equation}\label{Hdouble}
H=\frac{1}{2}(H_\textrm{I}+ e^{i\omega \Delta t}H_\textrm{II})
\end{equation}
At large delays $\Delta t\gg t_c$ the two corresponding modes are
distinguishable and Eq. \ref{Hmulti} still holds with $n_d$ replaced
by $2n_d$. This case is equivalent to doubling $t_c$ (see Ref.
\cite{Gisin}). When the delay is not large enough for
distinguishability, the two passes of Eq. \ref{Hdouble} interfere
and create oscillations between zero and an upper bound defined by
Eq. \ref{rate}. Therefore, in the single-mode case, a scan will
change the mode number between 2 and 1, thus the four-fold event
count would present a phase-independent dip. When more modes are
collected, the rate would oscillate between zero and a peak
envelope. The insert in Fig. \ref{fig2} is a fine scan that reveals
the expected oscillations.

Equation \ref{alpha} is a parameterization of Eq. \ref{Psi2multi}
for the case of $n_d=2$ and different weights, where
$\alpha=\frac{3}{16C}(c_1^4+c_2^4)$. Thus, the four-fold rate result
of Fig. \ref{fig2} is composed of a dip contribution from a single
mode per pass element and an oscillating contribution from a two
modes per pass element. Because the dip result of Fig. \ref{fig2}
was integrated over a long time, phase fluctuations from
experimental instability averaged the oscillating term. Considering
this model with the averaging effect, the best fit for the dip data
corresponds to $\alpha=0.80\pm 0.05$. The dashed curves in Fig.
\ref{fig2} demonstrate the predictions for different $\alpha$
values. The dip is sensitive to changes in $\alpha$ and transforms
into a peak for $\alpha\lesssim 0.64$.

\begin{figure}[tbp]
\includegraphics[angle=0,width=86mm]{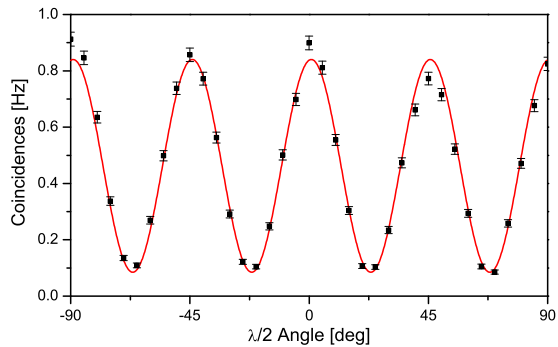}
\caption{\label{fig3}Four-fold (squares) visibility and its fit
(solid lines). Mode $a$ is detected in the $hv$ basis and mode $b$
is rotated between $hv$ and $pm$.}
\end{figure}

Great care has to be taken to ensure that the collected photons
originated from a single mode. Such measures include wavelength and
spatial filtering and compensation for temporal and spatial
birefringence walk-offs. Chirped pump pulses are also a source for
distinguishability as their duration is longer than their coherence
length\cite{Gisin}. Originally, it was suggested to measure $\alpha$
through the four-fold visibility, i.e. the contrast between
four-fold coincidence count rates at similar and orthogonal
polarization bases\cite{Tsujino}. Previously published values of
$\alpha$ are 37\% and 83\%, in Refs. \cite{Tsujino} and
\cite{Eisenberg2}, respectively. The results of the four-fold
visibility of our setup are presented in Fig. \ref{fig3}. From the
data we calculated $\alpha=86\%$. The agreement with the fit value
of $\alpha=80\pm5\%$ from the dip data (Fig. \ref{fig2}) and the
high $\alpha$ value, suggests the validity of our two modes per pass
analysis and the practical absence of PDC states with three or more
modes.

The pure $\psi_2^-$ content $\alpha$ is also bounded from below for
finite mode counts higher than two. The minimal value is
$\alpha_{min}=3/(2n_d+1)$. Thus, for example, a value of 37\%
indicates the necessary presence of at least four modes. As
visibility measurements are usually affected by many factors,
deriving $\alpha$ from the outcome is not sufficient for obtaining
the number of PDC modes collected in a specific setup. The sensitive
bunching measurement can help to remove this ambiguity.

\begin{figure}[tbp]
\includegraphics[angle=0,width=86mm]{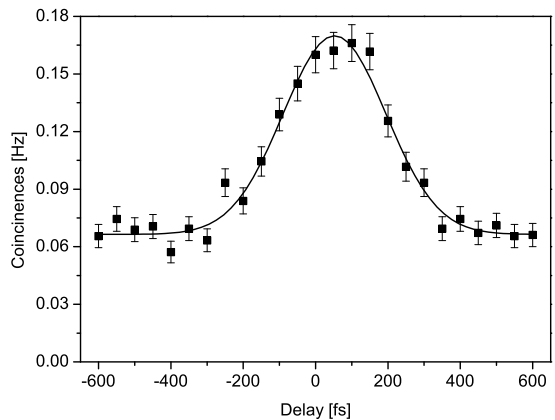}
\caption{\label{fig4} Four-fold coincidence rate as a function of
the delay $\Delta t$ between a $\psi^-$ and a $\phi^+$ states.
Polarization bases and all other parameters are the same as in Fig.
\ref{fig2}. Solid line is a Gaussian fit. Anti-bunching by a factor
of 2.6 is observed.}
\end{figure}

Bunching is not restricted to two $\psi_1^-$ states but can also
occur between any two identical Bell states. Furthermore, for 8 out
of the 10 possible combinations between two of the four Bell states,
bunching occurs. The remaining two combinations display another
interesting result -- Bell state anti-bunching. If we replace one of
the two $L^\dagger$ operators in Eq. \ref{Psi2double} by the
operator $K^\dagger$, defined as the creator of a $\phi^+$ state,
the last two minus signs of Eq. \ref{Psi4fold} change into plus
signs. As the delay $\Delta t$ approaches zero and the four elements
become indistinguishable, the four-fold count rate increases four
times through constructive interference, insensitive to the phase
delay as before. The results of such a measurement are presented in
Fig. \ref{fig4}. In Table \ref{table}, we summarize the results for
two polarization cases. For the third polarization option of $hvrl$
, anti-bunching occurs for $\psi^-\phi^-$ and for $\psi^+\phi^+$.

\begin{table}[tp]
\begin{tabular}{|c|cccc|}
\hline
(\textit{hv,pm}) & $\psi^-$ & $\psi^+$ & $\phi^-$ & $\phi^+$ \\
\hline
$\psi^-$ & B & B & B & A\\
$\psi^+$ & B & B & A & B\\
$\phi^-$ & B & A & B & B\\
$\phi^+$ & A & B & B & B\\
\hline
\end{tabular}
\begin{tabular}{|c|cccc|}
\hline
(\textit{pm,rl}) & $\psi^-$ & $\psi^+$ & $\phi^-$ & $\phi^+$ \\
\hline
$\psi^-$ & B & A & B & B\\
$\psi^+$ & A & B & B & B\\
$\phi^-$ & B & B & B & A\\
$\phi^+$ & B & B & A & B\\
\hline
\end{tabular}
\caption{\label{table}Bunching and anti-bunching of Bell states for
two polarization options of the modes ($a,b$). The letters B and A
mark bunching and anti-bunching combinations, respectively.}
\end{table}

Bell state bunching and anti-bunching are unique quantum optics
effects. They result from interference between two composite states
of photon pairs. Nevertheless, they are insensitive to the phase
between the pairs. In addition, the interference is perfect even if
the four down-conversion mode amplitudes are imbalanced. The Bell
state bunching effects are insensitive to the amplitudes of two
pairs from one pass (first and last terms of Eq. \ref{Psi2double}),
unlike any previous experiment.

In conclusion, we presented the bunching and anti-bunching of two
pairs of photons, each in a Bell state. Only the amplitudes of a
single pair from each of the two crystal passes of a pump pulse
contribute to this effect. It was shown how the bunching arises from
varying the number of distinguishable modes by adding a time delay
between the two pairs. The bunching contrast was related to the
$\psi_2^-$ content of the four photon state. Just like the HOM
effect plays a crucial role in projective measurements and
operations between single photons, the bunching of Bell states might
prove to be a useful tool in manipulating and projecting
multi-photon states.

\end{document}